\documentclass{aa}
\usepackage{graphicx,times,amssymb}
\usepackage[english]{babel}
\renewcommand{\d}{\mathrm{d}}

\def\be{\begin{equation}} \def\ee{\end{equation}}
\def\bea{\begin{eqnarray}} \def\eea{\end{eqnarray}}

\def\kg{{\bf k}}   

\begin{document}

\title{Improving the Accuracy of Cosmic Magnification Statistics}
\author{Brice M\'enard\inst{1,2}, Takashi Hamana\inst{3,1}, Matthias
  Bartelmann\inst{1}, Naoki Yoshida\inst{4,1}}
\institute{$^1$Max-Planck-Institut f\"ur Astrophysik, P.O.~Box 1317,
  D--85741 Garching, Germany\\
  $^2$Institut d'Astrophysique de Paris, 98 bis Bld Arago, F--75014,
  Paris, France\\
  $^3$National Astronomical Observatory of Japan, Mitaka, Tokyo
  181-8588, Japan\\
  $^4$Harvard-Smithsonian Center for Astrophysics, 60 Garden Street,
  Cambridge MA 02138, USA}

\date{\today}

\authorrunning{M\'enard et al.~}
\titlerunning{Improving the accuracy of Cosmic Magnification
  Statistics}

\abstract
 {The systematic magnification of background sources by the weak
  gravitational-lensing effects of foreground matter, also called
  \emph{cosmic magnification}, is becoming an efficient tool both for
  measuring cosmological parameters and for exploring the
  distribution of galaxies relative to the dark matter.  We extend
  here the formalism of magnification statistics by estimating the
  contribution of second-order terms in the Taylor expansion of the
  magnification and show that the effect of these terms was previously
  underestimated. We test our analytical predictions against numerical
  simulations and demonstrate that including second-order terms allows
  the accuracy of magnification-related statistics to be substantially
  improved. We also show, however, that both numerical and analytical
  estimates can provide only lower bounds to real correlation
  functions, even in the weak lensing regime. 
  We propose to use count-in-cells estimators rather than
  correlation functions for measuring cosmic magnification since they
  can more easily be related to correlations measured in numerical
  simulations.
}

\maketitle

\section{Introduction}

Gravitational lensing by large-scale structures magnifies sources and
distorts their images. The systematic distortion of faint background
galaxies near matter overdensities, the \emph{cosmic shear}, has been
measured by several groups in the past few years (Bacon et al.~2000,
2002; H\"ammerle et al.~2002; Hoekstra et al.~2002; Kaiser et
al.~2000; Maoli et al.~2001; R\'efr\'egier et al.~2002; Rhodes et
al.~2001; Van Waerbeke et al.~2000, 2001, 2002; Wittman et
al.~2000). It was found to be in remarkable agreement with theoretical
predictions based on the Cold Dark Matter model, and has already
provided new constraints on cosmological parameters (Van Waerbeke et
al.~2001).

In a similar way, systematic magnifications of background sources near
foreground matter overdensities, the \emph{cosmic magnification}, can
be measured and can provide largely independent constraints on
cosmological parameters (M\'enard \& Bartelmann 2002, M\'enard et
al.~2002).  Gravitational magnification has two effects: first, the
flux received from distant sources is increased, and the solid angle
in which they appear is stretched, thus their density is diluted. The
net result of these competing effects depends on how the loss of
sources due to dilution is balanced by the gain of sources due to flux
magnification. Sources with flat luminosity functions, like faint
galaxies, are depleted by cosmic magnification, while the number
density of sources with steep luminosity functions, like quasars, is
increased. Thus, cosmic magnification gives rise to apparent angular
cross-correlations between background sources and foreground matter
overdensities which are physically completely uncorrelated with the
sources. These overdensities can be traced by using the distribution
of foreground galaxies.

Numerous studies have confirmed the existence of quasar-galaxy
correlations on angular scales ranging from one arc minute to about
one degree, as expected from cosmic lensing (for a review, see
Bartelmann \& Schneider 2001; also Guimar\~aes et al.~2001). However,
the measured amplitude of these correlations has been systematically
higher than usually theoretically predicted.

While cosmic shear can directly be related to observable quantities
like image ellipticities, the theoretical interpretation of cosmic
magnification involves several approximations:

\begin{itemize}

\item the luminosity function of the sources is described by a
  power-law over the range probed by the flux limit of the
  observation; and

\item the magnification is assumed to fall into the weak lensing
  regime, i.e.~to deviate weakly from unity. Thus, the magnification
  can with sufficient accuracy be approximated by its first-order
  Taylor expansion and its deviation from unity becomes proportional
  to the lensing convergence alone.

\end{itemize}

While the first assumption is comfortably satisfied, in particular for
quasars, the validity of the second needs to be verified. This is the
goal of the present paper.

Our paper is structured as follows: first, we introduce the formalism
of the effective magnification and its Taylor expansion in
Sect.~\ref{formalism}. We then describe a number of statistics related
to the lensing convergence, and evaluate the amplitude of the
second-order terms which appear in the Taylor expansion. In
Sect.~\ref{section_simulation}, we describe the numerical simulations
we use to test our analytical results and estimate the accuracy of
several approximations for the magnification. As an application, we
investigate second-order effects on quasar-galaxy correlations in
Sect.~\ref{qso-gal}, and we summarise our results in
Sect.~\ref{conclusion}.

\section{Formalism\label{formalism}}

\subsection{Expanding the magnification}

Cosmic magnification can be measured statistically through
characteristic changes in the number density of the background
sources. Along a given line-of-sight, this effect depends on two
quantities:

\begin{itemize}

\item the magnification factor $\mu$, which describes whether sources
  are magnified or demagnified, depending on whether the matter along
  their lines-of-sight is preferentially over- or underdense compared
  to the mean,

\item and the logarithmic slope $\alpha$ of the source counts as a
  function of flux, which quantifies the amplitude of source
  number-count modifications due to flux magnification.  As mentioned
  in the introduction, magnification by gravitational lensing not only
  increases the observed flux, but also stretches the sky, thus the
  number density of sources on a magnified patch of the sky is
  reduced. The net magnification effect, called \emph{magnification
  bias}, depends on the balance between the number of sources lost by
  dilution and gained by flux magnification. The steeper the
  number-count function of the sources is, the more pronounced is the
  magnification bias.

\end{itemize}

If the number-count function of the background sources can be
described as a power law in a sufficiently wide range around the flux
limit of the observation, the magnification bias is quantified by the
\emph{effective magnification} $\mu^{\alpha-1}$. It directly expresses
the changes of the background source density caused by lensing through
the relation
\begin{equation}
  n(>S,\vec\theta)=\mu^{\alpha-1}(\vec\theta)\,n_0(>S)
\end{equation}
where $n_0(>S)$ is the intrinsic number-count function of sources
whose observed flux exceeds $S$ in the absence of lensing, and $n(>S)$
is the corresponding number-count function in presence of lensing.

The local properties of the gravitational lens mapping are
characterised by the convergence $\kappa$, which is proportional to
the surface mass density projected along the line-of-sight, and the
shear $\gamma$, which is a two-component quantity and describes the
gravitational tidal field of the lensing mass distribution. The
effective magnification is related to $\kappa$ and $\gamma$ through
\begin{equation}
  \mu^{\alpha-1}=\left[(1-\kappa)^2-|\gamma|^2\right]^{1-\alpha}\;,
\label{mu_def}
\end{equation}
where $|\gamma|=(\gamma_1^2+\gamma_2^2)^{1/2}$ is taken as the
absolute value of the shear. In the weak-lensing regime, both $\kappa$
and $|\gamma|$ are small compared to unity, and the previous
expression can be expanded in a Taylor series:
\begin{equation}
  \mu^{\alpha-1}=1+(\alpha-1)\left[
    2\kappa+(2\alpha-1)\kappa^2+|\gamma|^2
  \right]+\mathcal{O}(\kappa^3,|\gamma|^3)\;.
\label{mu_exp}
\end{equation}
Previous studies using analytical formulae for magnification
statistics focused only on the first-order term of this expansion,
i.e.~they used the approximation
$\mu^{\alpha-1}\approx1+2(\alpha-1)\,\kappa$, which potentially causes
the amplitude of the effect to be underestimated. In this section, we
investigate the second-order terms in the expansion and estimate their
contribution.

In doing so, we first note that $\kappa^2(\vec\theta)$ and
$|\gamma|^2(\vec\theta)$ share the same statistical properties
(e.g.~Blandford et al.~1991), because both $\kappa$ and $\gamma$ are
linear combinations of second-order derivatives of the lensing
potential. The identity of their statistics is most easily seen in
Fourier space. Since we will only deal with ensemble averages of the
magnification later on, $\kappa^2$ and $|\gamma|^2$ can be combined
into a single variable, which we denote by $\kappa$ for
simplicity. Thus, we can write for our purposes,
\begin{equation}
  \mu^{\alpha-1}=1+2(\alpha-1)\left[\kappa+\alpha\kappa^2\right]+
  \mathcal{O}(\kappa^3)\;.
\end{equation}
Observable effects are due to departures from the mean value of the
magnification. Therefore, the relevant quantity to correlate is
$\delta\mu^{\alpha-1}=\mu^{\alpha-1}-\langle \mu^{\alpha-1} \rangle$.
Then, up to second order in $\kappa^2$, the autocorrelation function
of the effective magnification is
\begin{eqnarray}
  \langle\delta\mu^{\alpha-1}(\vec\phi)\,
         \delta\mu^{\alpha-1}(\vec\phi+\vec\theta)\rangle &=&
  4(\alpha-1)^2\,\left[\langle\kappa(\vec\phi)
    \kappa(\vec\phi+\vec\theta)\rangle
    \right.\nonumber\\ &+&\left.
  2\alpha\langle\kappa(\vec\phi)
    \kappa^2(\vec\phi+\vec\theta)\rangle\right]\;,
\label{Taylor_2}
\end{eqnarray}
and the corresponding power spectrum can be expanded in a similar way,
\begin{equation}
  P_{\mu^{\alpha-1}}(s)=4(\alpha-1)^2\,\left[
  P_\kappa(s)+2\alpha P_{\mu,2}(s)\right]\;;
\end{equation}
the power spectrum $P_{\mu,2}(s)$ will be defined in Eq.~(\ref{pmu2})
below. The last two equations show that the importance of the
second-order terms in the expansion (\ref{mu_exp}) increases as the
number-count function of the background sources steepens, i.e.~as
$\alpha$ increases.  In the following, we will use $\alpha=2$ unless
stated otherwise. This value applies, for instance, to the number
counts of bright quasars with $m_\mathrm{B}<19.5$ (Pei 1995). For
simplicity, we abbreviate $\langle\delta\mu\,\delta\mu\rangle$ by
$\langle\mu\,\mu\rangle$.

\subsection{Second and Third-Order Correlations}

We will now estimate several $\kappa$-related statistical quantities
needed in the Taylor expansion of the magnification. For this purpose,
we first introduce the $\kappa$ projector such that
\begin{equation}
  \kappa(\vec\theta)=\int_0^{w_\mathrm{H}}\d
  w\,p_\kappa(w)\delta[\vec\theta f_K(w),w]
\end{equation}
can be written as a weighted line-of-sight projection of the density
contrast $\delta$ from the observer to the Hubble distance
$w_\mathrm{H}$. The projector is
\begin{eqnarray}
  p_\kappa(w)&=&\frac{3}{2}\,\Omega_0\left(\frac{H_0}{c}\right)^2
  \nonumber\\&\times&
  \int_w^{w_\mathrm{H}}\frac{\d w'}{a(w)}\,n_\mathrm{S}(w')\,
  \frac{f_K(w)\,f_K(w'-w)}{f_K(w')}\;,
\label{kappa_projector}
\end{eqnarray}
where $w$ is the radial coordinate distance, $f_K(w)$ is the comoving
angular-diameter distance, $n_\mathrm{S}(w)$ is the normalised
distance distribution of the sources, and $a(w)$ is the cosmological
scale factor. Using Limber's equation, we can then relate the
autocorrelation function of $\kappa$ to the dark-matter power spectrum
$P_\delta$,
\begin{eqnarray}
  \langle\kappa(\vec\phi)\kappa(\vec\phi+\vec\theta)\rangle&=&
  \int\d w~\frac{p^2_\kappa(w)}{f_K^2(w)} \nonumber\\ &\times&
  \int\frac{s\d s}{2\pi}P_\delta\left(\frac{s}{f_k(w)},w\right)
  \mathrm{J}_0(s\,\theta)\;,
\label{P_kappa_1}
\end{eqnarray}
where J$_0$ is the zeroth-order Bessel function, and the power
spectrum $P_\kappa$ corresponding to this correlation function is
\begin{equation}
  P_\kappa(s)=\int\d w~\frac{p^2_\kappa(w)}{f_K^2(w)}
  P_\delta\left(\frac{s}{f_k(w)},w\right)\;.
\end{equation}

As indicated by Eq.~(\ref{Taylor_2}), the estimation of second-order
terms requires the computation of the cross-correlation between
$\kappa$ and $\kappa^2$. We do this by first introducing a three-point
correlation function for $\kappa$ and then identifying two of its
three points. As usual, we define the three-point function by
\begin{equation}
  z_\kappa(\vec\theta_1,\vec\theta_2)=\langle
    \kappa(\vec\phi)\,
    \kappa(\vec\phi+\vec\theta_1)\,
    \kappa(\vec\phi+\vec\theta_2)\rangle\;.
\label{expression_triple}
\end{equation}
Using the $\kappa$ projector defined in (\ref{kappa_projector}), we
can then write
\begin{eqnarray}
  z_\kappa(\vec\theta_1,\vec\theta_2)&=&
  \int\d w_1\,p_\kappa(w_1)\int\d w_2\,p_\kappa(w_2)
  \int\d w_3\,p_\kappa(w_3)\nonumber\\&\times& 
  \left\langle
    \delta[f_K(w_1)\vec\phi,w_1]\,
    \delta[f_K(w_2)(\vec\phi+\vec\theta_1),w_2]
  \right.\nonumber\\&\times&\left.
    \delta[f_K(w_3)(\vec\phi+\vec\theta_2),w_3]
  \right\rangle\;.
\label{expression_z}
\end{eqnarray}
Next, we employ the approximation underlying Limber's equation, which
asserts that the coherence length of the density fluctuation field is
much smaller than the scales on which the projector $p_\kappa$ varies
appreciably.  Finally, we insert the expression for the bispectrum of
the dark-matter fluctuations detailed in Appendix~\ref{bispectrum},
and find
\begin{eqnarray}
  z_\kappa(\vec\theta_1,\vec\theta_2)&=&\int\d w\,p_\kappa^3(w)
  \int\frac{\d^2k_1}{(2\pi)^2}\,
  \mathrm{e}^{i\vec{k_1}\cdot\vec\theta_1\,f_K(w)}
  \int\frac{\d^2k_2}{(2\pi)^2}  \nonumber\\&\times&
  B_\delta(\vec k_1,\vec k_2,-\vec k_1-\vec k_2,w)\,
  \mathrm{e}^{i\vec{k}_2\cdot\vec\theta_2\,f_K(w)}\;,
\end{eqnarray}
where $B_\delta(\vec k_1,\vec k_2,\vec k_3)$ is defined by
\begin{equation}
  \langle
    \hat\delta(\vec k_1)\hat\delta(\vec k_2)\hat\delta(\vec k_3)
  \rangle=
  \delta_D(\vec k_1+\vec k_2+\vec k_3)\,
  B_\delta(\vec k_1,\vec k_2,\vec k_3)\;.
\end{equation}
Then, using Eq.~(\ref{expression_z}) and identifying two points of the
three-point correlation function
$\vec\theta_1\to\vec\theta_2\equiv\vec\theta$, we find
\begin{eqnarray}
  \langle\kappa(\vec\phi)\kappa^2(\vec\phi+\vec\theta)\rangle
  &=&\int\d w\,p_\kappa^3(w)\int\frac{\d^2k_1}{(2\pi)^2}\,
    \mathrm{e}^{i\vec k_1\cdot\vec\theta\,f_K(w)}
    \nonumber\\&\times&
  \int\frac{\d^2k_2}{(2\pi)^2}\,
  B_\delta(\vec k_1,\vec k_2,-\vec k_1-\vec k_2,w)\;.
\label{def_kkk}
\end{eqnarray}
The term $\langle\kappa(\vec\phi)\kappa^2(\vec\phi+\vec\theta)\rangle$
is a function of $\theta$ only. Its contribution $P_{\mu,2}(s)$ to the
power spectrum of the magnification is given by the inverse Fourier
transform of Eq.~(\ref{def_kkk}), which reads
\begin{eqnarray}
  P_{\mu,2}(s)&=&\int dw~\frac{p^3_\kappa(w)}{f_K^4(w)}
  \int\frac{d^2s'}{(2\pi)^2} \nonumber\\&\times&
  B_\delta\left(
    \frac{\vec s'}{f_k(w)},\frac{\vec s}{f_k(w)},
    \frac{-\vec s'-\vec s}{f_k(w)},w
  \right)\;.
\label{pmu2}
\end{eqnarray}

\subsection{Results and predictions}
\label{predictions}

We can now numerically evaluate the first two contributions to the
Taylor expansion of the magnification autocorrelation function defined
in Eq.~(\ref{Taylor_2}). As mentioned before, we use $\alpha=2$ here.

\begin{figure}[ht]
  \includegraphics[width=\hsize]{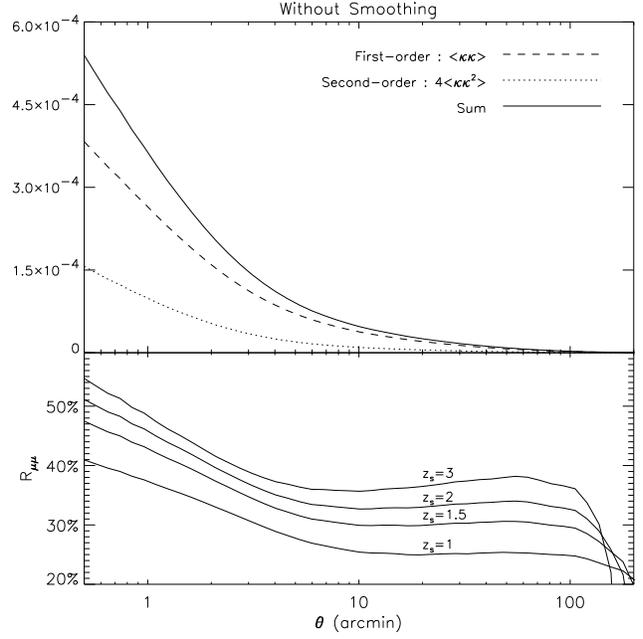}
\caption{The upper panel shows the amplitude of the two first terms of
  the Taylor expansion of the magnification autocorrelation, namely
  $\langle\kappa(\vec\phi)\kappa(\vec\phi+\vec\theta)\rangle$ (dashed
  line) and
  $4\langle\kappa(\vec\phi)\kappa^2(\vec\phi+\vec\theta)\rangle$
  (dotted line), using a source redshift of unity. The sum of these
  two terms is shown as the solid line. The lower panel details the
  relative contribution $R_{\mu\mu}$ of the second-order term for
  different source redshifts. The figure shows that the lowest-order
  approximation $\mu\approx1+2\kappa$ misses a substantial part of the
  amplitude of the magnification autocorrelation function. Given the
  accuracy of the bispectrum fitting formula, $R_{\mu\mu}$ is accurate
  to $\sim2\%$.}
\label{theoretical_kk}
\end{figure}

For evaluating the correlation functions, we use a CDM power spectrum
in a spatially flat Universe parameterised with $\Omega_0=0.3$,
$\sigma_8=0.9$, $h=0.7$ and $\Gamma=0.21$. The non-linear evolution of
the power spectrum and the bispectrum are computed according to the
formalisms developed by Peacock \& Dodds (1996) and Scoccimarro et
al.~(2000), see Appendix~\ref{bispectrum}. The upper panel of
Fig.~\ref{theoretical_kk} shows the first- and second-order
contributions (dashed and dotted lines, respectively) to the Taylor
expansion of the magnification for a fixed source redshift of
$z_\mathrm{s}=1$. The sum of the two contributions is shown by the
solid line. The figure shows that the contribution of the second-order
term reaches an amplitude of more than $30\%$ of the first-order term
on angular scales smaller than one arc minute. According to
Eq.~(\ref{Taylor_2}) which describes the Taylor expansion of the
magnification autocorrelation, we define the relative contribution of
the second-order compared to the first-order term as
\begin{equation}
  R_{\mu\mu}(\theta)=
  \frac{2\,\alpha\,\langle\kappa(\vec\phi)
                          \kappa^2(\vec\phi+\vec\theta)\rangle}
       {\langle\kappa(\vec\phi)\kappa(\vec\phi+\vec\theta)\rangle}\;.
\end{equation}
The lower panel shows this ratio in per cent for different source
redshifts as a function of angular scale. From the lower to the upper
curves, the source redshifts are $1$, $1.5$, $2$ and $3$. For each
source redshift, the contribution of the second term exhibits a
similar dependence on angular scale:

\begin{itemize}

\item on scales larger than a few degrees, the contribution drops to
  negligible values,

\item effects become relevant on smaller scales, with a fairly
  constant amplitude from a few degrees down to around 10 arc minutes,

\item on yet smaller scales, the second-order contribution increases
  steeply, due to the non-linear evolution of the density field.  For
  sources at redshift $2$, the amplitude of the second term reaches
  \emph{half} of the amplitude of the first term below one arc minute.

\end{itemize}

Thus, given the amplitude of $R_{\mu\mu}$, the correcting term
introduced in Eq.~(\ref{Taylor_2}) is relevant and must be taken into
account for describing the magnification autocorrelation with an
accuracy better than $30\%-50\%$ on scales smaller than a few degrees.

So far, we have only investigated the amplitude contributed by the
second-order term. In order to estimate the remaining contributions of
all missing terms of the magnification expansion, we will now use
numerical simulations allowing a direct computation of $\mu$ as a
function of the convergence $\kappa$ and the shear $\gamma$.

\section{Magnification statistics from numerical simulations}
\label{section_simulation}

\subsection{The ray-tracing simulation}

For testing the theoretical predictions we performed ray-tracing
experiments in a Very Large $N$-body Simulation (VLS) recently carried
out by the Virgo Consortium (Jenkins et al.~2001, and see also Yoshida
et al.~2001 for simulation details)%
\footnote{The ray-tracing data are available from T.~Hamana on
request, hamanatk@cc.nao.ac.jp} The simulation was performed using a
parallel P$^3$M code (MacFarland et al.~1998) with a force softening
length of $l_\mathrm{soft}\sim 30\,h^{-1}\mathrm{kpc}$. The simulation
employed $512^3$ CDM particles in a cubic box of
$479\,h^{-1}\mathrm{Mpc}$ on a side. It uses a flat cosmological model
with a matter density $\Omega_0=0.3$, a cosmological constant
$\Omega_\Lambda=0.7$, and a Hubble constant $h=0.7$. The initial
matter power spectrum was computed using CMBFAST (Seljak \&
Zaldarriaga 1996) assuming a baryonic matter density of
$\Omega_\mathrm{b}=0.04$. The particle mass
($m_\mathrm{part}=6.86\times 10^{10}h^{-1}M_\odot$) of the simulation
is sufficiently small to guarantee practically no discreteness effects
on dark-matter clustering on scales down to the softening length in
the redshift range of interest for our purposes (Hamana, Yoshida \&
Suto 2002).

The multiple-lens plane ray-tracing algorithm we used is detailed in
Hamana \& Mellier (2001; see also Bartelmann \& Schneider 1992 and
Jain, Seljak \& White 2000 for the theoretical basics); we thus
describe only aspects specific to the VLS $N$-body data in the
following. In order to generate the density field between $z=0$ and
$z\sim3$, we use a stack of ten snapshot outputs from two runs of the
$N$-body simulation, which differ only in the realisation of the
initial fluctuation field. Each cubic box is divided into 4 sub-boxes
of $479^2\times 119.75h^{-3}\mathrm{Mpc}^3$ with the shorter box side
being aligned with the line-of-sight direction. The $N$-body particles
in each sub-box are projected onto the plane perpendicular to the
shorter box side and thus to the line-of-sight direction. In this way,
the particle distribution between the observer and $z\sim3$ is
projected onto $38$ lens planes separated by
$119.75\,h^{-1}\mathrm{Mpc}$. Note that in order to minimise the
difference in redshift between a lens plane and an output of $N$-body
data, only one half of the outputs (i.e.~two sub-boxes) at $z=0$ are
used.

The particle distribution on each plane is converted into the surface
density field on either a $1024^2$ or $2048^2$ regular grid using the
triangular shaped cloud (TSC) assignment scheme (Hockney \& Eastwood
1988).  The two grid sizes are adopted for the following reasons:

\begin{itemize}

\item the $1024^2$ grid is chosen to maintain the resolution provided by
  the $N$-body simulation and removing at the same time the shot noise
  due to discreteness in the $N$-body simulation. Its computation
  follows the procedure described in Hamana \& Mellier (2001) and Jain
  et al.~(2000). The corresponding outputs will be labelled with
  \emph{large-scale smoothing} in the following.
\item the $2048^2$ grid is also chosen to examine effects of
  small-scale nonlinear structures which are smoothed in the
  \emph{large-scale smoothing} simulation. We should, however, note
  that in this case the shot noise is not sufficiently removed.
  Actually, the shot-noise power spectrum amplitude exceeds the
  convergence power spectrum on scales below $\sim1$~arcmin. In the
  following, therefore, we will only consider measured correlation
  functions on scales larger than 1~arcmin. The corresponding outputs
  will be labelled with \emph{small-scale smoothing} below.
\end{itemize}

Having produced surface density fields on all lens planes, $1024^2$
rays are traced backwards from the observer's point using the
multiple-lens plane algorithm (e.g.~Schneider, Ehlers \& Falco
1992). The initial ray directions are set on $1024^2$ grids with a
grid size of $0.25\,\mathrm{arcmin}$, thus the total area covered by
rays is $4.27^2\,$square degrees. We produced $36$ realizations of the
underlying density field by randomly shifting the simulation boxes in
the direction perpendicular to the line-of-sight using the periodic
boundary conditions of the $N$-body boxes. Note that the lens planes
coming from the same box are shifted in the same way in order to
maintain the clustering of matter in the box.

\begin{figure}[ht]
  \includegraphics[width=\hsize]{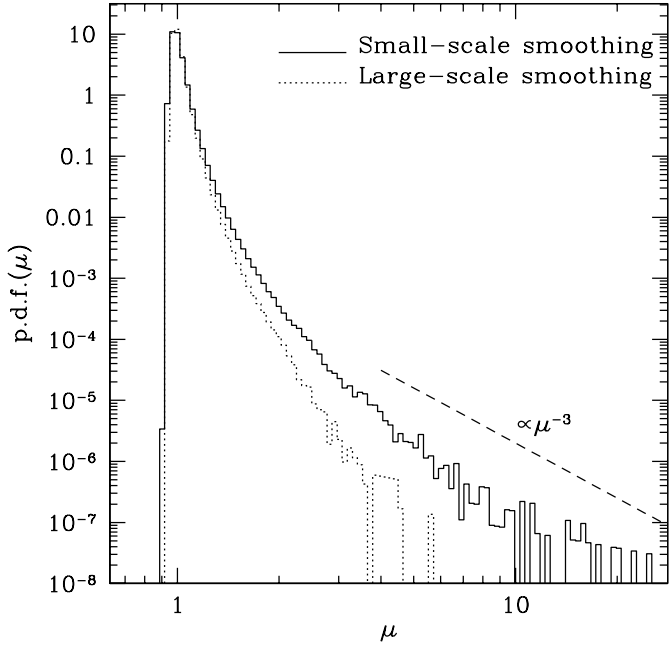}
\caption{Probability distribution of the magnification for our small-
  and large-scale smoothing simulations, assuming sources at redshift
  unity. The power law tail behaviour ($\propto\mu^{-3}$) found in
  the small-scale smoothing indicates the existence of caustics, while
  for large-scale smoothing, no prominent tail is shown which suggests
  caustics do not play a noticeable role.}
\label{pdf_mu}
\end{figure}

We point out that second and higher-order statistics of point-source
magnifications are generally ill-defined in presence of caustic curves
because the differential magnification probability distribution
asymptotically decreases as $\mu^{-3}$ for large $\mu$ (see
Fig.~\ref{pdf_mu}).  This is a generic feature of magnification near
caustics and is thus independent of the lens model. Strong lensing
effects on point sources near caustic curves give rise to rare, but
arbitrarily high magnification values in the simulations, and
therefore the variance of the measured statistics of $\mu$ cannot be
defined. However, the smoothing procedure introduced above allows this
problem to be removed because it smoothes out high density regions in
the dark matter distribution and thus the fractional area of high
magnification decreases.  In reality, infinite magnifications do not
occur, for two reasons. First, each astrophysical source is extended
and its magnification (given the surface brightness-weighted
point-source magnification across its solid angle) remains
finite. Second, even point sources would be magnified by a finite
value since for them, the geometrical-optics approximation fails near
critical curves and a wave-optics description leads to a finite
magnification (Schneider et al.~1992, Chap.~7).

\subsection{Filtering}
\label{filtering}

The computation of correlation functions from numerical simulations is
mainly affected by two effects; on large scales by the finite box size
of the dark matter simulation, and on small scales by the grid size
used for computing the surface density field from the particle
distribution. These boundaries set the limits for the validity of
correlation functions measured in numerical simulations. In other
words, this means that measuring a correlation function on a given
scale is relevant only if this scale falls within the range of scales
defined by the simulation. As shown in the previous section, our
method for computing the cross-correlation between $\kappa$ and
$\kappa^2$ consists of first computing a three-point correlation
function $\langle\kappa(\vec\phi)\kappa(\vec\phi+\vec\theta_1)
\kappa(\vec\phi+\vec\theta_2)\rangle$, and then identifying two of its
three points. In such a case, one of the correlation lengths of the
triple correlator becomes zero, thus necessarily smaller than the
smallest relevant scales in any simulation. This prevents us from
using any numerical simulation for directly comparing the results.

\begin{figure}[ht]
\begin{center}
  \includegraphics[width=\hsize]{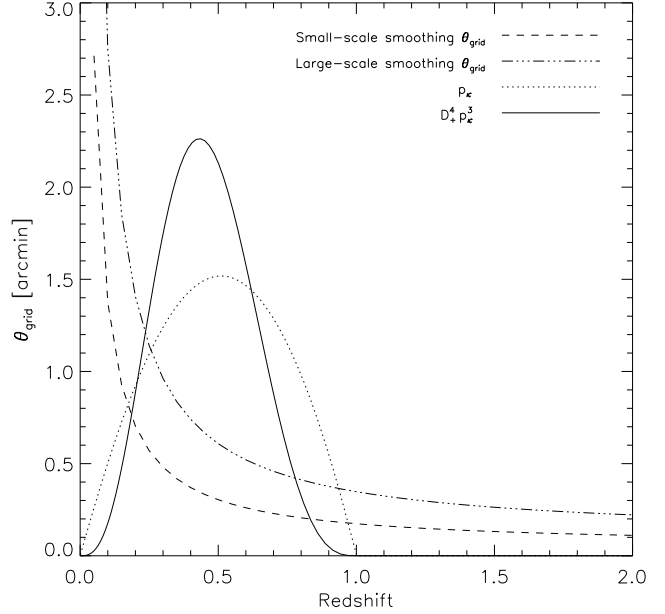}
\caption{Smoothing angle of the simulation as a function of redshift
  for the two ray-tracing schemes. In order to show the relevant
  quantities leading to the effective smoothing angle, we overplot the
  weighting function $W(z)=D_+^4(z)\,p_\kappa^3(z)$ (see
  Eqs.~[\ref{kappa_projector}] and [\ref{def_theta_s}]).}
\label{smoothing_z}
\end{center}
\end{figure}

In order to avoid this problem, and for comparing our analytical with
numerical results, we will introduce an effective smoothing into the
theoretical calculations, such that each value of $\kappa$ at a given
position $\vec\theta$ is evaluated by averaging the $\kappa$-values in
a disk of radius $\vec\theta_\mathrm{S}$ centred on $\vec\theta$.
Indeed, the limit imposed by the grid size of the simulation gives
rise to an unavoidable smoothing-like effect which cancels all
information coming from scales smaller than a corresponding smoothing
scale $\vec\theta_\mathrm{S}$. For this purpose, we introduce a
smoothed three-point correlator,
\begin{equation}
  z_\mathrm{smooth}(\vec\theta_2-\vec\theta_1,\vec\theta_3-\vec\theta_1)
  =\langle
    \kappa(\vec\theta_1)
    \kappa(\vec\theta_2)
    \kappa(\vec\theta_3)
  \rangle_{\theta_\mathrm{S}}
\end{equation}
\vspace{-.5cm}
\begin{eqnarray}
 ~~~~~~ &=& \int\d\vec\theta_1'\int\d\vec\theta_2'\int\d\vec\theta_3'
    \left\langle
      \kappa(\vec\theta_1')\kappa(\vec\theta_2')\kappa(\vec\theta_3')
    \right\rangle\nonumber\\&\times&
  W_{\theta_\mathrm{S}}(\vec\theta_1'-\vec\theta_1)
  W_{\theta_\mathrm{S}}(\vec\theta_2'-\vec\theta_2)
  W_{\theta_\mathrm{S}}(\vec\theta_3'-\vec\theta_3)\;,\nonumber
\end{eqnarray}
where the function $W_{\theta_\mathrm{S}}(\theta')$ is a normalised
top-hat window of radius $\theta_\mathrm{S}$. Introducing this
smoothing scheme into the expression for
$\langle\kappa(\vec\phi)\kappa^2(\vec\phi+\vec\theta)\rangle$ yields
\begin{eqnarray}
  z_\mathrm{smooth}(\theta_\mathrm{S})&=&
    \int\d w\frac{p^3_\kappa(w)}{f_K^4(w)}
    \int\frac{\d^2s_1}{(2\pi)^2}\int\frac{\d^2s_2}{(2\pi)^2}
  \nonumber\\&\times&
  \mathrm{I}(s_1\,\theta_\mathrm{S})\,
  \mathrm{I}(s_2\,\theta_\mathrm{S})\,
  \mathrm{I}(|\vec s_1+\vec s_2|\theta_\mathrm{S})
  \nonumber\\&\times&
  B_\delta\left(\frac{s_1}{f_k(w)},
                \frac{s_2}{f_k(w)},\frac{-s_1-s_2}{f_k(w)},w
          \right)\,e^{i\,\vec{s_1}\vec\theta}\;,
\end{eqnarray}
where $\mathrm{I}(x)=2\,\frac{\mathrm{J}_1(x)}{x}$. Similarly,
introducing the smoothing scheme into the two-point correlation
function gives
\begin{eqnarray}
  && w_\mathrm{smooth}(|\vec\theta_2-\vec\theta_1|)
  = \langle
    \kappa(\vec\theta_1)
    \kappa(\vec\theta_2)
     \rangle_{\theta_\mathrm{S}}
  \nonumber\\&=&
  \int\d w\frac{p^2_\kappa(w)}{f_K^2(w)}\int\frac{\d^2s}{(2\pi)^2}\,
  P\left(\frac{s}{f_K(w)},w\right)\nonumber\\&\times&
  |\mathrm{I}(s\,\theta_\mathrm{S})|^2\,e^{i\,\vec{s}\vec\theta}\;.
\label{smoothed_w_kk}
\end{eqnarray}

The effective smoothing scale depends on two parameters:

\begin{itemize}

\item the evolution of the apparent grid size of the simulation as a
  function of redshift, and

\item the radial selection function of the dark-matter field whose
  correlation function has to be measured.

\end{itemize}

These quantities are plotted in Fig.~\ref{smoothing_z}. In order to
use a unique smoothing scale valid on the final convergence map, we
define the effective angular smoothing scale by
\begin{equation}
  \theta_\mathrm{S}=\int\d z\,W(z)\,\theta_\mathrm{grid}(z)\;,
\label{def_theta_s}
\end{equation}
where $W(z)$ is the relevant normalised selection function along the
line-of-sight.  Measuring $w_\mathrm{s}$ means probing the power
spectrum along the line-of-sight, weighted by
$p_\kappa^2(z)$. Therefore, we will use
$W(z)=D_+^2(z)\,p_\kappa^2(z)$, where $D_+(z)$ is the growth
factor. In a similar way, we will use $W(z)=D_+^4(z)\,p_\kappa^3(z)$
for measuring $z_\mathrm{smooth}$. The numerical values of the
corresponding effective angles are presented in Table~\ref{my_table}.
\begin{table}[ht]
\begin{center}
\begin{tabular}{ccc}
\hline
  & small-scale smoothing & large-scale smoothing \\
\hline
  $w_\mathrm{smooth}$ &
  $\theta_\mathrm{S}=0.40$ & $\theta_\mathrm{S}=0.80$ \\
  $z_\mathrm{smooth}$ &
  $\theta_\mathrm{S}=0.39$ & $\theta_\mathrm{S}=0.78$ \\
\hline
\end{tabular}
\end{center}
\caption{Effective smoothing angles in arc minutes for $w_s$ and $z_s$
  computed from Eq. \ref{def_theta_s} as a function of simulation
  resolution.}
\vspace{-.5cm}
\label{my_table}
\end{table}

The second important difference between analytical calculations and
measurements in numerical simulations is the finite box size
effect. Indeed, the analytical correlation functions presented above
were computed taking into account all modes in the power
spectrum. However, the finite size of the box used in the simulation
introduces an artificial cutoff in the power spectrum since
wavelengths larger than the box size are not sampled by the
simulation. This effect can also be taken into account in the
analytical calculations by simply cancelling all the power on
wavelengths with wave number $k<k_\mathrm{min}$. The boxes we use have
a comoving size of $480\,h^{-1}\mathrm{Mpc}$ which corresponds to
$k_\mathrm{min}=0.013\,h\,\mathrm{Mpc}^{-1}$.

\subsection{Comparing $\langle\kappa\,\kappa\rangle$ and $\langle
  \kappa\,\kappa^2 \rangle$}

With the help of the filtering schemes introduced in the previous
section, we can now compare our theoretical predictions with
correlation functions measured from the numerical simulations. We
first compare the amplitude and angular variation of the two first
terms of the Taylor expansion of the magnification separately. In the
next section, we will then compare their sum to the total
magnification fully computed from the simulation.

\begin{figure}[ht]
\begin{center}
  \includegraphics[width=\hsize]{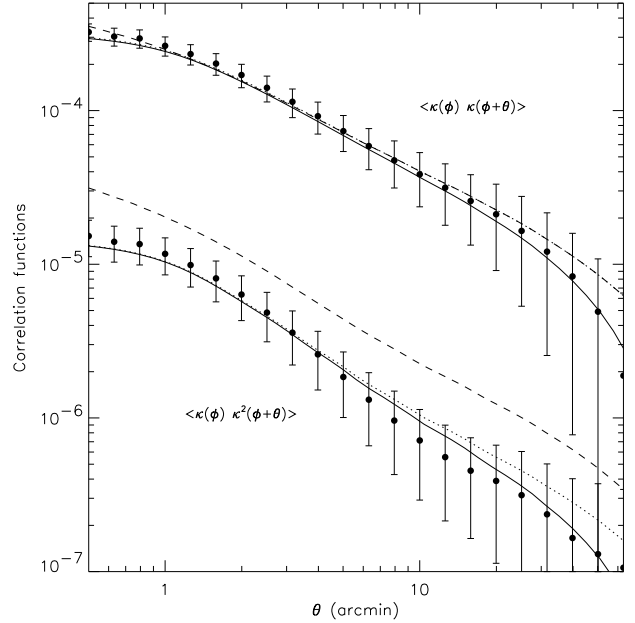}\hfill
\caption{Comparison between theoretical predictions and measurements
  from numerical simulations assuming sources at redshift unity.  The
  upper and lower curves show $\langle\kappa\kappa\rangle$ and
  $\langle\kappa\kappa^2\rangle$, respectively. The points are
  measurements from the large-scale smoothing simulations, with the
  error bars showing the variance among $36$ different
  realisations. The dotted lines show the analytical computations
  taking into account the smoothing scale of the simulation.  The
  solid lines additionally include a cut in the power spectrum for
  cancelling the wavelengths not covered by the simulation.  The
  dashed line presents the same statistics without any
  smoothing. Obviously, the smoothing effects are crucial for the
  $\langle\kappa\kappa^2\rangle$ cross-correlation.  }
  \label{numerical_kk}
\end{center}
\end{figure}

In Fig.~\ref{numerical_kk}, we overplot analytical and numerical
results. The upper curve shows the autocorrelation function of
$\kappa$ as a function of angular scale. We plot in circles the
average measurement from $36$ realisations of the simulation, and the
corresponding 1-$\sigma$ error bars to show the accuracy of the
numerical results as a function of angular scale. The solid line shows
the analytical prediction, including effective smoothing and an
artificial cut of the power at scales below $k_\mathrm{min}$. The
agreement is good on all scales. For comparison, the dotted line shows
the result if we do not impose the large-wavelength cut, and the
dashed line is the result if no cut and no smoothing are applied. In
both cases, the deviations from the fully filtered calculation remain
small since we are probing angular scales within the range allowed by
the simulation.

\begin{figure*}[ht]
  \includegraphics[width=.45\hsize]{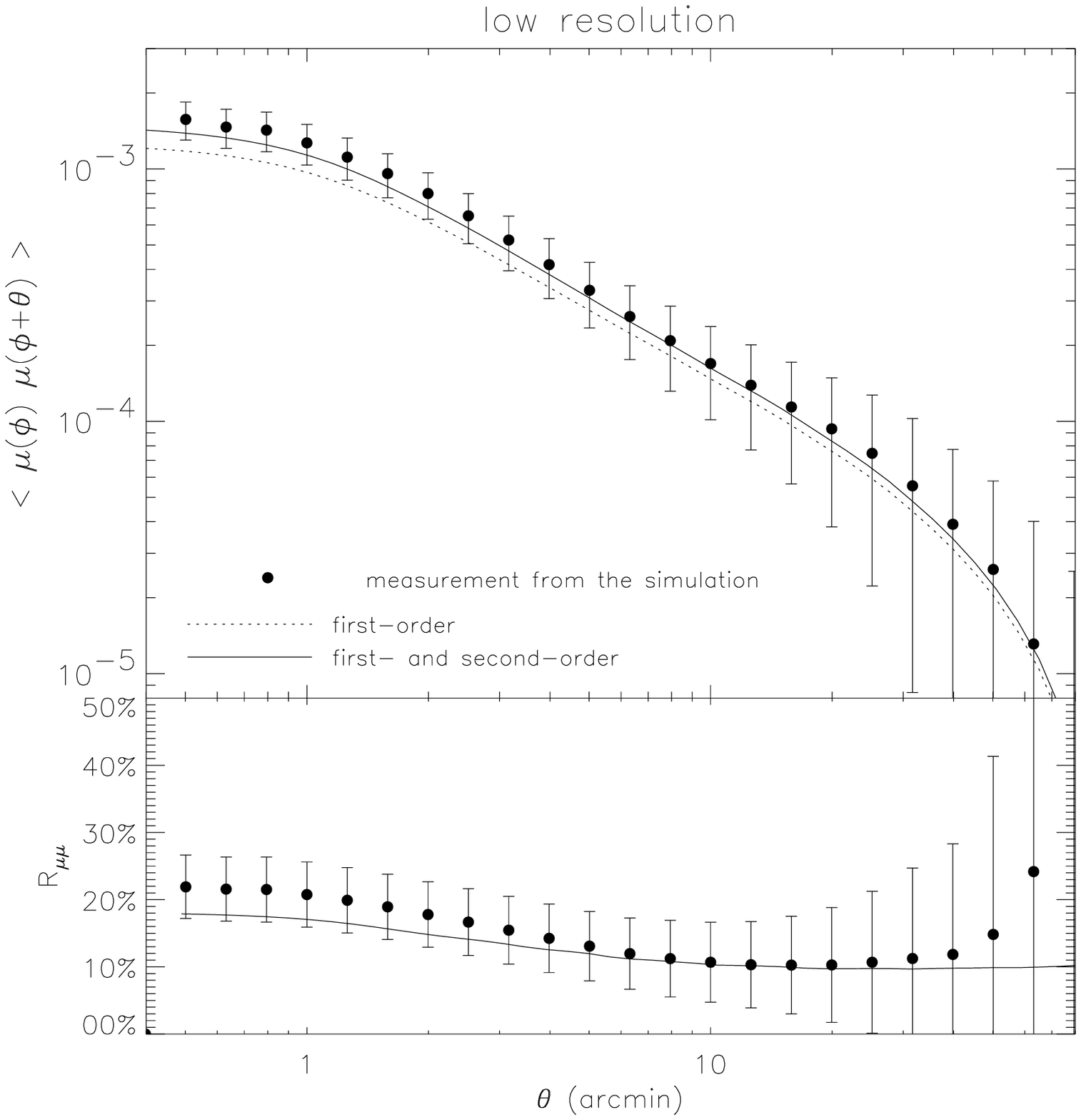}
  \hfill
  \includegraphics[width=.45\hsize]{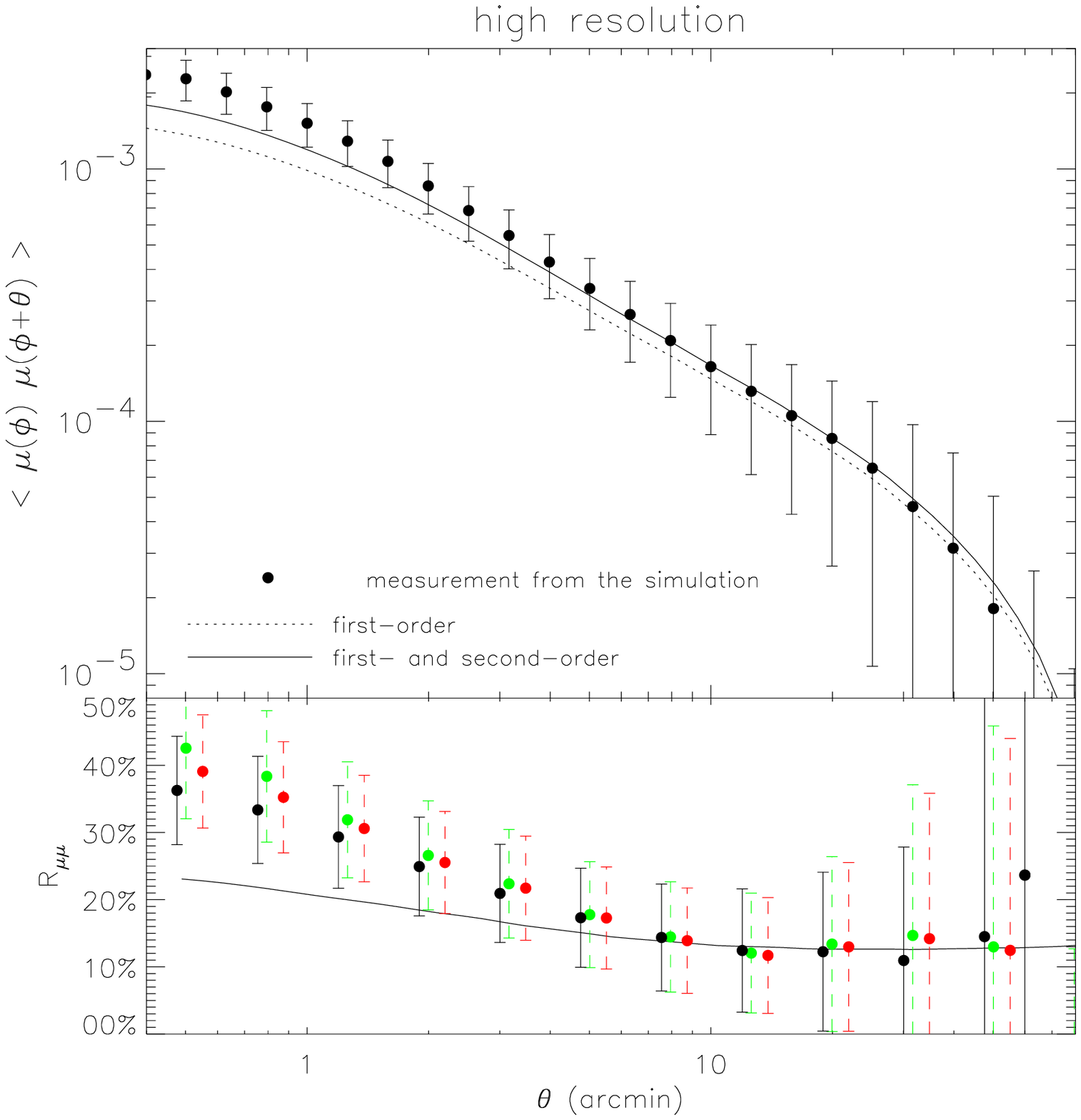}
\caption{The upper panel shows a comparison between the magnification
  autocorrelation measured from the simulation compared to the
  analytical estimation for a source redshift of unity.  The circles
  show averaged measurements from $36$ realisations of the simulation
  and the corresponding 1-$\sigma$ error bars. The dashed line shows
  the analytical estimation using the approximation
  $\mu\approx1+2\kappa$. The solid line shows the improvement given by
  the second-order term of the Taylor expansion of $\mu$. The lower
  panel shows the relative contribution $R_{\mu\mu}$, both measured
  from the simulation and estimated using expansion terms of $\mu$ up
  to second order. In the right panel, each angular point shows three
  different measurements, taking into account the complete
  magnification maps (solid-line error bars), or maps where pixels
  were masked where the magnification value exceeds 8 or 4
  (dashed-line error bars, from left to right).}
\label{check_2_terms}
\end{figure*}

The lower curves in Fig.~\ref{numerical_kk} show a quantity
proportional to the second-order correction of the Taylor expansion,
namely the correlation function $\langle\kappa\,\kappa^2\rangle$. In
the same way as before, the circles show average measurements from
$36$ realisations, and the error bars denote the corresponding
1-$\sigma$ deviation. The prediction including smoothing and
small-wavelength cut (solid line) shows a relatively good agreement
given the expected accuracy of the bispectrum fitting formula, which
is approximately 15\% (Scoccimarro \& Couchman 2000). This time,
including smoothing changes the amplitude dramatically, and this
effect affects all scales (see the dashed line). As discussed before,
this is expected since we are measuring a three-point correlator on
triangles which have one side length smaller than the angular grid
size of the simulation. Finally, as shown by the difference between
the dotted and solid lines, cancelling the power on scales where
$k<k_\mathrm{min}$ again improves the agreement on large scales.

The agreement between our analytical and numerical computations of
$\langle\kappa\,\kappa\rangle$ and $\langle\kappa\,\kappa^2\rangle$
demonstrates the validity of the formalism introduced in
Sect.~\ref{formalism} as well as the choice of the effective smoothing
scale (Eq.~[\ref{def_theta_s}]) for describing the second-order term
in the Taylor expansion of the magnification.

\subsection{Deviations from simulated $\mu$-statistics}

We now want to investigate how well the second-order expansion
describes the full magnification expression (\ref{mu_def}) which can
be computed using maps of $\kappa$, $\vec\gamma$ and $\omega$ (a net
rotation term which arises from lens-lens coupling and the lensing
deflection of the light ray path; see Van Waerbeke et al.~2001b)
obtained from the simulations (see Hamana et al.~2000 for more
detail).

Before doing so, we recall that the amplitude of the magnification
autocorrelation measured from the simulation depends on the smoothing
scale, as seen in Sect.~\ref{filtering}, since $\mu$ is nonlinear in
the density field. Therefore, all the following comparisons are valid
for a given effective smoothing length only.

We further emphasise that two problems will complicate this
comparison. First, our analytical treatment is valid in the
weak-lensing regime only, i.e.~as long as convergence and shear are
small compared to unity, $\kappa\ll1$, $|\gamma|\ll1$. While most
light rays traced through the numerical simulations are indeed weakly
lensed, a non-negligible fraction of them will experience
magnifications well above two, say. Such events are restricted to
small areas with high overdensities and thus affect the magnification
statistics only at small angular scales. Second, a separate problem
sets in if and where caustics are formed. The magnification of light
rays going through caustics is infinite, and the magnification
probability distribution near caustics drops like $\mu^{-3}$ for
$\mu\to\infty$. As noted above, second- or higher-order statistics of
$\mu$ then become meaningless because they diverge.

Departures of the numerical from the analytical results will thus have
two distinct reasons, viz.~the occurrence of non-weak magnifications
which causes the analytical to underestimate the numerical results on
small angular scales; and the formation of caustics, which causes
second-order magnification statistics to break down entirely. Both
effects will be demonstrated below. They can be controlled or
suppressed in numerical simulations by smoothing, which makes lensing
weaker, or by masking highly magnified light rays or regions
containing caustics.

In Fig.~\ref{check_2_terms}, we plot with circles the autocorrelation
function $\langle\mu(\vec\phi)\,\mu(\vec\phi+\vec\theta)\rangle$
measured from the large- and small-scale smoothing simulations in the
left and right panels, respectively. The presence of caustics is more
pronounced in the case of small-scale smoothing than in the
large-scale smoothing simulations. The dotted line shows the
theoretical prediction given by the first-order term of the Taylor
expansion, namely
$4\,\langle\kappa(\vec\phi)\kappa(\vec\phi+\vec\theta)\rangle$. This
yields a low estimate of the correlation, with a discrepancy of order
10\% on large scales, and more than 20\% below a few arc minutes.

As expected from the preceding discussion, this level of discrepancy
also depends on the effective smoothing scale and can increase if
simulations with a smaller grid size are used. Estimating the
contribution of the two lowest-order terms of $\mu^{\alpha-1}$, we
computed in Sect.~\ref{predictions} a lower bound to this discrepancy
for a real case without smoothing, and found it to reach a level of
25\% at large scales, and above 30\% below a few arc minutes. The
smoothed results taking the additional contribution of the
second-order term into account are plotted as solid lines, and give a
much better agreement, as expected. To quantify this in more detail,
the lower panels of the figure show several contributions compared to
the first-order term, i.e.~to $4\langle\kappa\kappa\rangle$.

\begin{itemize}

\item The symbols show the additional amplitude of the magnification
  statistics measured from the simulation, compared to the first-order
  term also obtained from the simulation,
\begin{equation}
  R_{\mu\mu}^\mathrm{num}=
  \frac{\langle\mu(\vec\phi)\mu(\vec\phi+\vec\theta)
        \rangle_\mathrm{num}}
       {4\,\langle\kappa(\vec\phi)\kappa(\vec\phi+\vec\theta)
        \rangle_\mathrm{num}}-1\;.
\end{equation}
  The error bars indicate the 1-$\sigma$ deviation across $36$
  realisations.
\item The solid line shows the contribution of the second-order
  relative to the first-order term computed from the analytical
  expression including the effective smoothing,
\begin{equation}
  R_{\mu\mu}^\mathrm{smooth}=\frac{2\alpha\langle
    \kappa(\vec\phi)\kappa^2(\vec\phi+\vec\theta)
    \rangle_{\theta_\mathrm{S}}}
    {\langle\kappa(\vec\phi)\kappa(\vec\phi+\vec\theta)
    \rangle_{\theta_\mathrm{S}}}
\end{equation}
  with $\alpha=2$.

\end{itemize}

In each case, we use the appropriate reference for $\langle
\kappa(\vec\phi)\kappa(\vec\phi+\vec\theta)\rangle$, i.e.~the
numerical measurement in the first and the analytical estimation in
the second case. Indeed, the measurement of
$\langle\kappa\,\kappa\rangle$ from the simulation agrees with the
analytical estimation within some uncertainty, which is due to
numerical effects like the finite number of dark-matter boxes used for
simulating the light cone. It introduces a bias into our comparisons
which is impossible to separate from the real offset due to all
higher-order terms of the Taylor expansion that were not taken into
account. The two contributions plotted in Fig.~\ref{check_2_terms} are
thus of different nature, but are suitable for a relative comparison.

As the lower panel of the large-scale smoothing simulation shows, the
simple $4\,\langle\kappa\,\kappa\rangle$ estimate of the magnification
misses 20\% of the real amplitude near one arc minute. This
discrepancy almost vanishes after adding the contribution of the
second-order term, which gives at all scales a final agreement on the
per cent level: the additional amplitude reaches 19\% at the smallest
scales of the figure, compared to a value of 20\% given by the
simulation, and agrees within better than one per cent on larger
scales. Therefore, taking into account the
$2\,\alpha\,\langle\kappa\kappa^2\rangle$ correction allows the
accuracy to be increased by a factor of $\sim20$ compared to the
approximation 4$\langle\kappa\kappa\rangle$, in the case of our
\emph{large-scale smoothing} simulation. On the largest scales,
between $6$ and $30$ arc minutes, the agreement even improves. Above
these scales, the numerical results do not allow any relevant
comparison because the number of available independent samplings
corresponding to a given separation decreases. On scales below a few
arc minutes, the offset between the measured points and the analytical
estimate gives the amplitude of all higher-order terms neglected in
the Taylor expansion of the magnification. As we can see, their
contribution is on the one per cent level for the large-scale
smoothing simulation.

The curves shown in the right panel demonstrate how the use of a
smaller smoothing scale increases the discrepancy between the
analytical and the numerical results. The fraction of non-weakly
magnified light rays increases, and caustics appear which give rise to
a power-law tail in the magnification probability distribution. We
investigate the impact of the rare highly magnified light rays by
masking pixels where the simulated magnification exceeds 4 or 8, and
show that caustics have no noticeable effect on the amplitude of the
magnification autocorrelation function determined from these simulated
data. Note, however, that the impact of the caustics depends on the
source redshift. The higher the redshift, the more caustics appear,
and the larger is their impact on the correlation amplitude.

Imposing lower masking thresholds removes a significant fraction of
the area covered by the simulation, changing the spatial magnification
pattern and thus the magnification autocorrelation function. The
corresponding measurements are represented by the dashed error bars in
the lower right panel of Fig.~\ref{check_2_terms}. We note that the
error bars of $R_{\mu\mu}^\mathrm{num}$ computed with the small-scale
smoothing simulation become larger at small scales compared to the
lower left panel. This reflects the fact that second-order
magnification statistics are ill-defined once caustics appear. In the
next section, we will investigate similar smoothing effects on
cross-correlations between magnification and dark matter
fluctuations. These quantities are not affected by problems of poor
definition when the smoothing scale becomes small, and therefore do
not show larger error bars at small scales when the smoothing scale
decreases.

These comparisons show that the approximation $\mu\approx 1+2\kappa$
misses a non-negligible part of the total amplitude of weak-lensing
magnification statistics. The formalism introduced in
Sect.~\ref{formalism} allows second-order corrections to be described
with or without smoothing of the density field. This provides a better
description of the correlation functions, but still gives a lower
amplitude than the simulation results.  As we noticed, the analytic
computation based on the Taylor expansion is sufficiently accurate
only in the weak lensing regime. In reality, however, the strong
lensing, which can not be taken into account in the analytic
formalism, has a significant impact on the magnification correlation
especially at small scales as shown in the small-scale smoothing
simulation. Therefore, one should carefully take the strong lensing
effect into consideration when one interprets the magnification
related correlation functions. However, we will see in the next
section that counts-in-cells estimators are less affected by the
strong lensing than correlation functions and thus enable better
comparisons of observations with results from simulations.

\section{Applications to quasar-galaxy correlations}
\label{qso-gal}

As a direct application of the formalism introduced previously, we now
investigate the effects of second-order terms on a well-known
magnification-induced correlation, namely the quasar-galaxy
cross-correlation (the results can also be applied to galaxy-galaxy
correlations induced by magnification; Moessner \& Jain 1998). In
order to estimate cosmological parameters from this kind of
correlations, we then suggest the use of a more suitable estimator
using counts-in-cells rather than two-point correlation functions. It
has the advantage of making the observational results more easily
reconciled with the ones from numerical simulations.

\subsection{Formalism and correcting terms}

The magnification bias of large-scale structures, combined with galaxy
biasing, leads to a cross-correlation of distant quasars with
foreground galaxies. The existence of this cross-correlation has
firmly been established (e.g.~Ben{\'\i}tez \& Mart{\'\i}nez-Gonz\'alez
1995; Williams \& Irwin 1998; Norman \& Impey 1999; Norman \& Williams
2000; Ben{\'\i}tez et al.~2001; Norman \& Impey 2001). M\'enard \&
Bartelmann (2002) showed that the Sloan Digital Sky Survey (York et
al.~2000) will allow this correlation function to be measured with a
high accuracy. Its amplitude and angular shape contain information on
cosmological parameters and the galaxy bias factor. Thus, it is
important to accurately describe these magnification-related
statistics in order to avoid a biased estimation of cosmological
parameters as well as the amplitude of the galaxy bias.

As shown in Bartelmann (1995), the lensing-induced cross-correlation
function between quasars and galaxies can be written as
\begin{eqnarray}
  w_\mathrm{QG}(\theta)&\equiv&
  \langle\delta_\mathrm{QSO}(\vec\phi)\,
         \delta_\mathrm{gal}(\vec\phi+\vec\theta)
  \rangle\nonumber\\ &=&
  \langle\delta\mu^{\alpha-1}(\vec\phi)\,
         \delta_\mathrm{gal}(\vec\phi+\vec\theta)\rangle\;.
\end{eqnarray}
Using the above formalism, we can expand the effective magnification
fluctuation $\delta\mu^{\alpha-1}$ up to second order and find the
correcting term:
\begin{equation}
  w_\mathrm{QG}(\theta)=2\,(\alpha-1)\,\left[
    \langle\kappa\delta_\mathrm{gal}\rangle
    +\alpha\,\langle\kappa^2\delta_\mathrm{gal}\rangle
  \right]\;.
\end{equation}
The second term is proportional to $\alpha$ (contrary to the factor
$2\alpha$ in Eq.~[\ref{Taylor_2}]), since there is only one
contribution of the magnification. Therefore, the expected effects
will be roughly half of those on the autocorrelation of the effective
magnification seen in the previous section. Assuming a linear bias $b$
between galaxies and dark matter, the cross-correlation between
$\delta_\mathrm{gal}$ and $\kappa^2$ can be written as
\begin{eqnarray}
  &&\langle
    \delta_\mathrm{gal}(\vec\phi)\kappa^2(\vec\phi+\vec\theta)
  \rangle=b\,\langle
    \delta_{DM}(\vec\phi)\,\kappa^2(\vec\phi+\vec\theta)
  \rangle\nonumber\\
  &=&\int\d w~\frac{p^2_\kappa(w)\,p_\delta(w)}{f_K^4(w)}
     \int\frac{\d^2s_1}{(2\pi)^2}\int\frac{\d^2s_2}{(2\pi)^2}
  \nonumber\\&\times&
  B_\delta\left(
    \frac{s_1}{f_k(w)},\frac{s_2}{f_k(w)},\frac{-s_1-s_2}{f_k(w)},w
  \right)\,\mathrm{e}^{i\vec{s_1}\vec\theta}\;,
\label{def_dk2}
\end{eqnarray}
where $p_\delta(w)$ is the normalised distance distribution of the
galaxies. For this example, we will use
\begin{equation}
  p_\delta(z)\,\d z=\frac{\beta\,z^2}{z_0^3\,\Gamma(3/\beta)}\,
  \exp\left[-\left(\frac{z}{z_0}\right)^\beta\right]\,\d z\;,
\end{equation}
with $\beta=1.5$ and $z_0=0.3$.

\begin{figure}[ht]
  \includegraphics[width=\hsize]{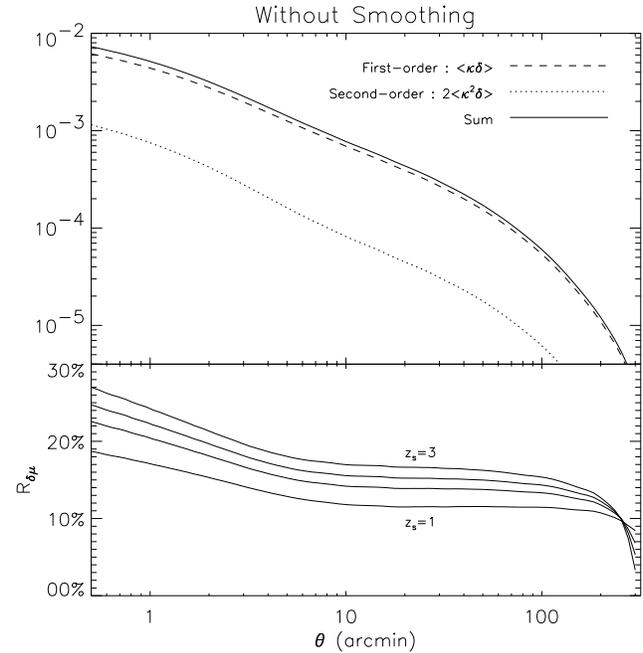}
\caption{The upper panel shows the amplitude of the normalised
  quasar-galaxy correlation $w_\mathrm{QG}/2(\alpha-1)$ as a function
  of angular separation. We show the first two terms of the Taylor
  expansion of this correlation, namely $\langle
  \delta(\vec\phi)\kappa(\vec\phi+\vec\theta)\rangle$ (dashed line)
  and $2\langle\delta(\vec\phi)\kappa^2(\vec\phi+\vec\theta)\rangle$
  (dotted line), using a source redshift of unity. The sum of these
  two terms is shown as the solid line. The lower panel details the
  relative contribution $R_{\delta\mu}$ and of the second-order term
  for different source redshifts, namely $z=1$, $1.5$, $2$ and $3$
  from bottom to top.}
\label{plot_w_qg}
\end{figure}

The results are shown in Fig.~\ref{plot_w_qg}. As we can see, previous
estimates using the approximation $\mu\approx 2\,\kappa$ missed
approximately $15\%$ of the amplitude on small scales for quasars at
redshift unity. Using quasars at redshift $2$, these effects reach up
to 25\%. These offsets, which are only lower limits, would lead to
biased estimates of $\Omega_0$ or $b$, for example.

\begin{figure}[ht]
  \includegraphics[width=\hsize]{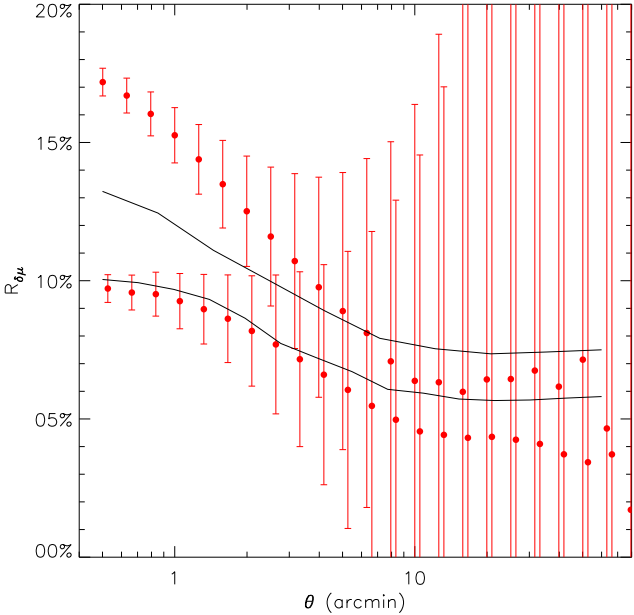}
\caption{Comparison of $R_{\delta\mu}$ given by the theoretical
  calculation and measured from the numerical simulation. The results
  are shown for the large- and small-scale smoothing simulations from
  bottom to top. They show that a second-order description of the
  $\langle\mu\delta\rangle$ cross-correlation gives good results when
  the smoothing is large, but misses some part of the total amplitude
  in the case of our small-scale smoothing ray-tracing.}
\label{plot_R_mu_delta}
\end{figure}

As for the magnification autocorrelation, we can compare our
theoretical estimates against numerical estimations. We can first introduce
a coefficient $R_{\delta\mu}$ describing the accuracy of our second-order 
correction:
\begin{equation}
  R_{\delta\mu}(\theta)=
  \frac{\alpha\,\langle\delta(\vec\phi)
                          \kappa^2(\vec\phi+\vec\theta)\rangle}
       {\langle\delta(\vec\phi)\kappa(\vec\phi+\vec\theta)\rangle}\;.
\end{equation}
We plot the results in Fig.~\ref{plot_R_mu_delta}. Note that contrary
to the magnification autocorrelation, this quantity does not suffer
from poor definition, even without smoothing.  The difference can be
seen by the same size of the error bars between the two simulation
results at small scales, whereas they were larger in the case of
$\langle \mu\mu \rangle$ for the small-scale smoothing simulation
(Fig.~\ref{check_2_terms}).  The results for $R_{\delta\mu}$ are very
similar those obtained for $R_{\mu\mu}$: for the large-scale smoothing
ray-tracing we find very good agreement which reaches the one percent
level on small scales.  However, when the smoothing length decreases,
we see from the small-scale smoothing outputs that we are missing a
part of the total amplitude on small scales, which shows that
higher-order terms play a non negligible role on those scales.

\subsection{Using count-in-cells estimators}

For precisely estimating cosmological parameters as well as the
amplitude of the galaxy bias, it is necessary to employ theoretical
magnification statistics that closely describe the observables.
However, we have seen in Sect.~\ref{section_simulation} that
analytical estimates as well as numerical simulations have intrinsic
limitations and prevent us from accurately describing usual $n$-point
correlation functions related to magnification statistics.

Besides, it is possible to focus on another estimator closely related
to correlation functions, namely a count-in-cells estimator, which
naturally smoothes effects originating from the density field and can
thus more easily be reconciled with numerical simulations. So far,
quasar-galaxy or galaxy-galaxy correlations have been quantified
measuring the excess of background-foreground pairs at a given angular
separation. Instead, we can correlate the amplitude of the background
and foreground fluctuations, both measured inside a given aperture. We
will therefore introduce a count-in-cells estimator,
\begin{eqnarray}
  \bar w_\mathrm{QG}(\theta)&=&
  \left\langle
    \delta_\mathrm{QSO}(\vec\phi)\,
    \delta_\mathrm{gal}(\vec\phi)
  \right\rangle_\theta\nonumber\\&&
  \left\langle
    \delta\mu^{\alpha-1}(\vec\phi)\,
    \delta_\mathrm{gal}(\vec\phi) 
  \right\rangle_\theta\;,
\end{eqnarray}
where the subscript $\theta$ indicates averaging of
$\delta_\mathrm{QSO}(\phi)$ and $\delta_\mathrm{gal}(\phi)$ inside a
cell of radius $\theta$. In practice, this estimator is intended to be
applied to galaxy-galaxy rather than to quasar-galaxy correlations,
since the average angular separation between bright distant quasars is
of order one degree for current surveys, thus averaging the source
counts inside cells with radii of several arc minutes will not be
relevant. Using galaxies as background sources, this limitation occurs
at much smaller scales.

Using a first-order Taylor expansion for the magnification, the new
estimator $\bar w_{\rm QG}(\theta)$ can be written
\begin{eqnarray}
  \frac{\bar w_\mathrm{QG}(\theta)}{2\,(\alpha-1)\,b}&=&
  \int\d w\frac{p_\kappa(w)p_\delta(w)}{f_K^2(w)}
  \int\frac{s\d s}{2\pi}\nonumber\\&\times&
  P\left(\frac{s}{f_k(w)},w\right)\,|\mathrm{I}(s\,\theta)|^2\;,
\end{eqnarray}
where $\mathrm{I}(x)=2\,\frac{\mathrm{J}_1(x)}{x}$. This expression
differs from the 2-point correlation function (\ref{P_kappa_1}) by its
Fourier-space filtering of the power spectrum. The additional
smoothing wipes out the power on scales smaller than the physical
scale corresponding to the angular smoothing scale $\theta$. For any
observational result to be compared to a numerical simulation,
$\theta$ and the smoothing scale used in the simulation will have to
be carefully adapted to each other and to the redshift distribution of
the foreground galaxy distribution.

Note that gravitational lensing by the foreground galaxies themselves
is entirely irrelevant here. The angular scale on which galaxies act
as efficient lenses is on the order of one arc second and below, much
smaller than the angular scales we are concerned with. Moreover, the
probability for a quasar to be strongly lensed by a galaxy is well
below one per cent. Bartelmann \& Schneider (1991) demonstrated this
point explicitly by including galaxies into their numerical
simulations and showing they had no noticeable effect.

\section{Conclusion}
\label{conclusion}

As surveys mapping the large-scale structure of the Universe become
wider and deeper, measuring cosmological parameters as well as the
galaxy bias with cosmic magnification will become increasingly
efficient and reliable. Therefore, an accurate theoretical
quantification of magnification statistics becomes increasingly
important.

Previous estimates of cosmic magnification relied on the assumption
that the magnification deviates sufficiently little from unity that it
can be accurately approximated by its first-order Taylor expansion
about unity, i.e.~$\mu\approx1+2\,\kappa$. In this paper, we have
tested the validity of this assumption in the framework of
magnification statistics, by investigating the second-order terms in
the Taylor expansion of $\mu$. We have shown that:

\begin{itemize}

\item Second-order terms can be related to the cross-correlation
  between $\kappa$ and $\kappa^2$,

\item their importance increases as the number-count function of the
  background sources steepens, i.e.~as $\alpha$ increases,

\item their amplitude is \emph{not} negligible: for the magnification
  autocorrelation, their contribution is typically on the order of
  30\%-50\% at scales below one degree. Therefore, previous estimates
  of cosmic magnification were systematically low.

\end{itemize}

For testing our theoretical calculations, we have compared our results
to magnification statistics found in numerical simulations by
performing ray-tracing experiments in a very large $N$-body
simulation.  We have first checked the validity of our formalism
describing the correlation $\langle \kappa\kappa^2 \rangle$, and
demonstrated the importance of including an effective smoothing into
the analytical calculations. Indeed, $\mu$ is nonlinear in the density
field and the amplitude of magnification statistics measured from
numerical simulations depends therefore on the available resolution.

Using a simulation with an effective smoothing scale of 0.8~arc
minutes, we found that our second-order formalism is accurate to the
percent level for describing magnification autocorrelations. Compared
to previous estimates, this improves the accuracy by a factor of
$\approx20$. For smaller effective smoothing scales, the contribution
of third- and higher-order terms becomes important on scales below a
few arc minutes.

Finally we have applied our formalism to observed correlations, like
quasar-galaxy and galaxy-galaxy correlations due to lensing. We have
shown that second-order corrections increase their amplitude by 15\%
to 25\% on scales below one degree. These correlations are valuable
tools to probe cosmological parameters as well as the galaxy
bias. However, even including our correcting terms, analytical or
numerical estimates of magnification statistics can only provide lower
bounds to the real amplitude of the correlation functions in the
weak-lensing regime. Thus, we propose using count-in-cells estimators
rather than correlation functions since the intrinsic smoothing in
determining counts-in-cells allows the observational results to be more
directly related to those obtained in numerical simulations.

\section*{Acknowledgements}

We thank Francis Bernardeau and St\'ephane Colombi for helpful
discussions. This work was supported in part by the TMR Network
``Gravitational Lensing: New Constraints on Cosmology and the
Distribution of Dark Matter'' of the EC under contract
No. ERBFMRX-CT97-0172.

\appendix

\section{Bispectrum and Non-Linear evolution}
\label{bispectrum}

The bispectrum can be estimated using second-order perturbation
theory. Indeed, an expansion of the density field to second nonlinear
order as
\begin{equation}
  \delta(\vec x)\approx\delta^{(1)}(\vec x)+\delta^{(2)}(\vec x)\;,
\end{equation}
where $\delta^{(2)}$ is of order $(\delta^{(1)})^2$ and represents
departures from Gaussian behaviour, yields the bispectrum
\begin{eqnarray}
  \langle\delta_1\delta_2\delta_3\rangle&\simeq&
  \langle\delta_1^{(1)}\delta_2^{(1)}\delta_3^{(1)}\rangle+
  \langle\delta_1^{(1)}\delta_2^{(1)}\delta_3^{(2)}\rangle
  \nonumber\\&+&
  \mbox{cyclic terms (231, 312)}\;.
\label{developpement}
\end{eqnarray}
The first term in Eq.~(\ref{developpement}) vanishes because the
density fluctuation field is Gaussian to first order, hence the third
moment of $\delta^{(1)}$ is zero. Thus, the leading term in
Eq.~(\ref{developpement}) is of the order of
$\langle\delta_1^{(1)}\delta_2^{(1)}\delta_3^{(2)}\rangle$ and can be
quantified using second-order perturbation theory.

The bispectrum $B_\delta(\kg_1,\kg_2,\kg_3)$ is defined only for closed
triangles formed by the wave vectors $\kg_1,\kg_2,\kg_3$. It can be
expressed as a function of the second-order kernel $F(\kg_1,\kg_2)$
and the power spectrum
\begin{eqnarray}
  B_\delta(\kg_1,\kg_2,\kg_3)&=&2~F(\kg_1,\kg_2)~P(k_1)P(k_2)\nonumber\\
  &+&2~F(\kg_2,\kg_3)~P(k_2)P(k_3)\nonumber\\
  &+&2~F(\kg_1,\kg_3)~P(k_1)P(k_3)\;.
\label{bispectredef}
\end{eqnarray}
For describing the bispectrum on all scales, we use the fitting
formula derived by Scoccimarro \& Couchman (2000) for the non-linear
evolution of the bispectrum in numerical simulations of CDM models,
extending previous work for scale-free initial conditions.  In that
case, we have
\begin{eqnarray}
  F(\kg_1,\kg_2)&=&{5\over 7}~a(n,k_1)a(n,k_2)\nonumber\\
  &+&{1\over 2}{\kg_1\cdot\kg_2\over k_1 k_2}\left({k_1\over k_2}+
  {k_2\over k_1}\right)~b(n,k_1)b(n,k_2)\nonumber\\ &+&{2\over
  7}\left({\kg_1\cdot \kg_2\over k_1 k_2}\right)^2~c(n,k_1)c(n,k_2),
\label{F_expression}
\end{eqnarray}
with
\begin{eqnarray}
  a(n,k)&=&{1+\sigma_8^{-0.2}(z)\left[0.7~Q_3(n)\right]^{1/2}
  (q/4)^{n+3.5}\over 1+(q/4)^{n+3.5}} \cr
  b(n,k)&=&{1+0.4~(n+3)~q^{n+3}\over 1+q^{n+3.5}}\cr
  c(n,k)&=&{1+4.5/\left[1.5+(n+3)^4\right](2q)^{n+3}\over
  1+(2q)^{n+3.5}},
\end{eqnarray}
and $q \equiv k/k_\mathrm{NL}(z)$, where $4\pi
k_\mathrm{NL}^3P_L(k_\mathrm{NL})=1$, and $P_\mathrm{L}(k)$ is the
linear power spectrum at the desired redshift. The effective spectral
index is taken from the linear power spectrum as well. The function
$Q_3(n)$ is given by
\begin{equation}
  Q_3(n)={(4-2^n)\over (1+2^{n+1})},
\label{Q3def}
\end{equation}
For more detail, see Scoccimarro \& Couchman (2000).

\end{document}